\documentstyle[amssymb,epsfig,secnumtab]{fbssuppl}

\title{Virtual Compton Scattering---Generalized Polarizabilities
of Nucleons and Pions}
\author{S. Scherer\thanks{This work was supported by the 
Deutsche Forschungsgemeinschaft 
(SFB 201) and a grant from NATO.}
}
\institute{Institut f\"{u}r Kernphysik, Johannes Gutenberg-Universit\"at,
J. J. Becher Weg 45, D-55099 Mainz, Germany}

\begin{document}

\maketitle
\begin{abstract}
   Virtual Compton scattering off nucleons and pions at low
energies is discussed.
   Predictions for the generalized polarizabilities of the nucleon 
are presented within the framework of heavy-baryon chiral perturbation
theory and the linear sigma model.
   First results for the generalized polarizabilities of the charged
pion in chiral perturbation theory at ${\cal O}(p^4)$ are shown.
\end{abstract}

\section{Introduction}
   Real Compton scattering (RCS) off a stable particle has a long history 
as a means to study the dynamics responsible for the internal structure 
of a system.
   In this context, low-energy theorems (LET) play an important role in 
defining a reference point for the accuracy which has to 
be achieved in order to obtain information beyond global properties of the
target and, thus, to potentially distinguish between different models.
   The famous low-energy theorem of Low \cite{t:ss:Low_1954} and  
Gell-Mann and Goldberger \cite{t:ss:GellMann_1954} for Compton scattering
of real photons off a nucleon is based on the requirements of gauge 
invariance, Lorentz covariance, crossing symmetry, and the discrete 
symmetries.
   The low-energy amplitude is specified up to and including terms linear
in the photon energy.
   The Taylor series coefficients are expressed in terms of the 
charge, mass, and magnetic moment of the target.
   Terms of second order in the frequency, which are not determined by
this LET, can be parametrized in terms of two new structure
constants, the electric and magnetic polarizabilities $\alpha$ and 
$\beta$ (see, e.g., 
\cite{t:ss:Holstein_1990,t:ss:Lvov_1993}).
   
   Recently, the investigation of low-energy virtual Compton scattering (VCS) 
as tested in, e.g., the reactions $e^-p \to e^-p \gamma$ 
\cite{t:ss:vcsexp,t:ss:d'Hose}  and 
$\pi^- e^-\to\pi^- e^-\gamma$ \cite{t:ss:Moinester}, 
has attracted a lot of interest.
   The possibilities to investigate the structure of the target increase
substantially, if virtual photons are used since (a) photon energy and momentum
can be varied independently and (b) longitudinal components of the 
transition current are accessible. 
   For the nucleon, the model-independent properties of the low-energy VCS 
amplitude have been identified in \cite{t:ss:Guichon_1995,t:ss:Scherer_1996}
whereas the spin-zero case has been discussed in Ref.\ 
\cite{t:ss:Fearing_1998}.
   In \cite{t:ss:Guichon_1995} the model-dependent part beyond the LET was 
analyzed in terms of a multipole expansion.  
   Keeping only terms linear in the energy of the final photon, 
the corresponding amplitude was parametrized in terms of 
ten so-called generalized polarizabilities (GPs) which are functions
of the three-momentum transfer of the virtual photon in the VCS process.
   The number of independent GPs reduces to six, if 
charge-conjugation invariance is imposed 
\cite{t:ss:Drechsel_1997,t:ss:Drechsel_1998}.
   Predictions for the generalized polarizabilities of the nucleon
have been obtained in various frameworks \cite{t:ss:Guichon_1995,%
t:ss:Liu_1996,%
t:ss:Vanderhaeghen_1996,t:ss:Metz_1996,t:ss:Hemmert_1997a,%
t:ss:Kim_1997,t:ss:Pasquini_1998}
(for an overview, see Ref.\ \cite{t:ss:Guichon_1998}).

\section{Kinematics and LET}
\begin{figure}[ht]
\begin{center}
\epsfig{file=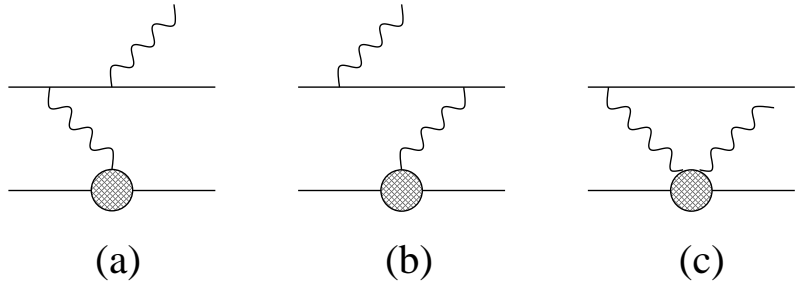,width=300pt}
\caption{Bethe-Heitler diagrams (a) and (b). VCS diagram (c).
The four-momenta of the virtual photons in the BH diagrams and the
VCS diagram differ from each other.
\label{fig:diagrams}}
\end{center}
\end{figure}

   The invariant amplitude consists of a Bethe-Heitler piece,
where the real photon is emitted by the initial or final electrons,
and the VCS contribution (see Fig.\ \ref{fig:diagrams}),
\begin{equation}
{\cal M}={\cal M}_{\mbox{\footnotesize BH}}+{\cal M}_{\mbox{\footnotesize 
VCS}}.
\end{equation}

\noindent   
In the following we will be concerned with the invariant amplitude for VCS, 
\begin{equation}
\label{mvcs}
{\cal M}_{\mbox{\footnotesize VCS}}
=-ie^2\epsilon_\mu \epsilon_\nu'^\ast M^{\mu\nu}
=-ie^2\epsilon_\mu M^\mu
=ie^2\left(\vec{\epsilon}_T\cdot\vec{M}_T
+\frac{q^2}{\omega^2}\epsilon_z M_z\right),
\end{equation}
where $\epsilon_\mu=e\bar{u}\gamma_\mu u/q^2$ 
is the polarization vector of the virtual photon
($e>0,e^2/4\pi \approx 1/137$), and where use of 
current conservation has been made.
   In the center-of-mass system, using the Coulomb gauge for the final
real photon, the transverse (longitudinal) part of ${\cal M}_{VCS}$ for
the nucleon can be expressed in terms of eight (four) independent structures
\cite{t:ss:Scherer_1996,t:ss:Hemmert_1997a}, 
\begin{equation}
\vec{\epsilon}_T\cdot \vec{M}_T=\vec{\epsilon}\,'^\ast \cdot 
\vec{\epsilon}_T A_1 + \cdots,\quad
M_z=\vec{\epsilon}\,'^\ast \cdot \hat{q} A_9 + \cdots,
\end{equation}
   where the functions $A_i$ depend on three kinematical variables,
$|\vec{q}|$, $\omega'=|\vec{q}\,'|$, $z=\hat{q}\cdot\hat{q}\,'$.
   For the spin zero case only one longitudinal and two transverse
structures are required \cite{t:ss:Fearing_1998,t:ss:Drechsel_1997}.

   Extending the method of Gell-Mann and Goldberger 
\cite{t:ss:GellMann_1954} to VCS, model-independent predictions for the 
functions $A_i$ were obtained in \cite{t:ss:Scherer_1996}.
   For example, the result for $A_1$ up to second order in $|\vec{q}|$
and $\omega'$ is
\begin{eqnarray}
\label{a1}
A_1&=&-\frac{1}{M}+\frac{z}{M^2}|\vec{q}|
-\left(\frac{1}{8M^3}+\frac{r^2_E}{6M}-\frac{\kappa}{4M^3}
-\frac{4\pi\alpha}{e^2}\right)\omega'^2\nonumber\\
&&+\left(\frac{1}{8M^3}+\frac{r^2_E}{6M}-\frac{z^2}{M^3}
+\frac{(1+\kappa)\kappa}{4M^3}\right)|\vec{q}|^2.
\end{eqnarray}
   To this order, all $A_i$ can be expressed in terms of 
$M$, $\kappa$, $G_E$, $G_M$, $r^2_E$, $\alpha$, and $\beta$.
   For $|\vec{q}|=\omega'$, Eq.\ (\ref{a1}) reduces to the well-known
RCS result.   

   In Ref.\ \cite{t:ss:Fearing_1998} the low-energy behavior of
the VCS amplitude of $\pi^-(p_i)+\gamma^\ast(\epsilon,q)\to
\pi^-(p_f)+\gamma(\epsilon',q')$ was found to be of the form
\begin{equation}
\label{mvcspion}
{\cal M}_{\mbox{\footnotesize VCS}}=-2ie^2 F(q^2)\left[
\frac{p_f\cdot\epsilon'^\ast (2 p_i+q)\cdot\epsilon}{s-m_\pi^2}
+\frac{p_i\cdot\epsilon'^\ast (2 p_f-q)\cdot\epsilon}{u-m_\pi^2}
-\epsilon\cdot\epsilon'^\ast\right]+{\cal M}_R,
\end{equation}
where $F(q^2)$ is the electromagnetic form factor of the pion,
$s=(p_i+q)^2$, and $u=(p_i-q')^2$.
   The residual term ${\cal M}_R$ is separately gauge invariant and
at least of second order, i.e.\ ${\cal O}(qq,qq',q'q')$.

\section{Generalized Polarizabilities}

   For the purpose of analyzing the model-dependent terms 
specific to VCS, the invariant amplitude 
${\cal M}_{\mbox{\footnotesize VCS}}$ is split 
into a (generalized) pole piece ${\cal M}_A$ and a residual part 
${\cal M}_B$.  
   For the nucleon,  the s- and u-channel pole diagrams are calculated using 
electromagnetic vertices of the form
\begin{equation} \label{f1f2vertex}
\Gamma^\mu(p',p)=\gamma^\mu F_1(q^2)+i\frac{\sigma^{\mu\nu}q_\nu}{2M} F_2(q^2),
\,q=p'-p, \end{equation}
where $F_1$ and $F_2$ are the Dirac and Pauli form factors, 
respectively.
   The corresponding amplitude ${\cal M}^{\gamma^\ast\gamma}_A$ contains all 
irregular terms as $q\to 0$ or $q'\to 0$ and is separately gauge invariant
\cite{t:ss:Guichon_1995,t:ss:Scherer_1996}.
   For the pion, the situation is somewhat more complicated due to
the fact that even for real photons the s- and u-channel pole diagrams are 
not separately gauge-invariant.
   A natural starting point is given by Eq.\ (\ref{mvcspion}) with
${\cal M}_B\equiv{\cal M}_R$.

   The generalized polarizabilities in VCS \cite{t:ss:Guichon_1995} result 
from an analysis of ${\cal M}_{B}^{\gamma^{\ast} \gamma}$ in terms of 
electromagnetic multipoles $H^{(\rho' L', \rho L)S}(\omega' , |\vec{q}|)$,  
where $\rho \, (\rho')$ denotes the type of the initial 
(final) photon ($\rho = 0$: charge, C; $\rho = 1$: magnetic, M;
$\rho = 2$: electric, E).  
   The initial (final) orbital angular momentum is denoted by 
$L \, (L')$, and $S$ distinguishes between non-spin-flip $(S = 0)$ and 
spin-flip $(S = 1)$ transitions.
   For the pion, only the case $S=0$ applies.
   According to the LET for VCS, ${\cal M}^{\gamma^\ast\gamma}_B$ is at 
least linear in the energy of the real photon.
   A restriction to the lowest-order, i.e.\ linear terms in $\omega'$ leads 
to only electric and magnetic dipole radiation in the final state.
   Parity and angular-momentum selection rules (see Table 
\ref{tab:mulan}) then allow for 3 scalar 
multipoles $(S = 0)$ and 7 vector multipoles $(S = 1)$. 
\begin{table}[hbt]
\caption{Multipolarities of initial and final states}
\label{tab:mulan}
\begin{tabular}
{|c|c|c|}
\hline
$J^P$&final real photon&initial virtual photon\\
\hline
$\frac{1}{2}^-$&E1&C1,E1\\
\hline
$\frac{3}{2}^-$&E1&C1,E1,M2\\
\hline
\hline
$\frac{1}{2}^+$&M1&C0,M1\\
\hline
$\frac{3}{2}^+$&M1&C2,E2,M1\\
\hline
\end{tabular}
\end{table}
   The corresponding ten GPs, $P^{(01,01)0}$, ...,
${\hat{P}}^{(11,2)1}\,$,
are functions of $|\vec{q}|^2$, 
   where mixed-type polarizabilities, 
${\hat{P}}^{(\rho' L' , L)S} (|\vec{q}|^2)$, have been introduced which are 
neither purely electric nor purely Coulomb type (see 
\cite{t:ss:Guichon_1995}
for details). 
   Only six of the above ten GPs are independent, if charge-conjugation 
symmetry is imposed \cite{t:ss:Drechsel_1997,t:ss:Drechsel_1998}. 
   For example, for a charged pion, the constraint for the Compton
tensor reads \cite{t:ss:Fearing_1998,t:ss:Drechsel_1997}
\begin{equation}
{\cal M}_{\pi^+}^{\mu\nu}(p_f,q';p_i,q)
\stackrel{C}{=}{\cal M}_{\pi^-}^{\mu\nu}(p_f,q';p_i,q)
\stackrel{crossing}{=}{\cal M}_{\pi^+}^{\mu\nu}(-p_i,q';-p_f,q),
\end{equation}
   generating one relation between originally three GPs
\cite{t:ss:Drechsel_1997,t:ss:Metz_1996}.
      
   Relations between the GPs at $|\vec{q}|=0$ and the four 
spin-dependent RCS polarizabilities $\gamma_i$ of Ref.\ 
\cite{t:ss:Ragusa_1993} were discussed in \cite{t:ss:Drechsel_1998}: 
\begin{equation}
\label{rpgp}
\gamma_3 =  - \frac{e^2}{4 \pi}
\frac{3}{\sqrt{2}} P^{(01,12)1} (0),\quad 
\gamma_2 + \gamma_4 
=  - \frac{e^2}{4 \pi} \frac{3 \sqrt{3}}{2 \sqrt{2}} P^{(11,02)1}(0),
\end{equation}
i.e., only two of the four $\gamma_i$ can be related to GPs at $|\vec{q}|=0$.

\section{Generalized Polarizabilities of the Nucleon}
   In the following we will discuss the predictions for the generalized
polarizabilities of the nucleon obtained within the heavy-baryon formulation 
of chiral perturbation theory (HBChPT) and the linear sigma model.

   In Ref.\ \cite{t:ss:Hemmert_1997a} the VCS amplitude
was calculated using HBChPT to third order in the external momenta.
   At ${\cal O}(p^3)$, contributions to the GPs are generated by nine one-loop 
diagrams and the $\pi^0$-exchange $t$-channel pole graph (see
\cite{t:ss:Hemmert_1997a}).   
   For the loop diagrams only the leading-order Lagrangians are
required,
\begin{equation}
\label{lpin1pipi2}
\widehat{\cal L}_{\pi N}^{(1)}=\bar{N}_v (i v \cdot D + g_A S \cdot u) N_v, 
\quad 
{\cal{L}}_{\pi \pi}^{(2)} = 
\frac{F_{\pi}^2}{4} 
\mbox{Tr}\left[ \nabla_{\mu} U (\nabla^{\mu} U)^{\dagger}\right],
\end{equation}
   where $N_v$ represents a non-relativistic nucleon field, and 
$U = {\mathrm{exp}}(i \vec \tau \cdot \vec \pi/F_{\pi})$ contains the pion 
field. 
   The covariant derivatives $\nabla_{\mu}U$ and $D_{\mu} N_v$ include
the coupling to the electromagnetic field, 
   and $u_\mu$ contains in addition the derivative coupling of a pion.
   In the heavy-baryon formulation the spin matrix is given by
$S^{\mu} = i \gamma_5 \sigma^{\mu\nu} v_{\nu}$, where $v^\mu$ is a
four-vector satisfying $v^2=1, v_0\ge1$.
   Finally, for the $\pi^0$-exchange diagram one requires in addition to
Eq.\ (\ref{lpin1pipi2}) the $\pi^0\gamma\gamma^\ast$ vertex 
provided by the Wess-Zumino-Witten Lagrangian,
\begin{equation}
\label{wzwpi0}
{\cal{L}}_{\gamma\gamma\pi^0}^{(WZW)} =  -\frac{e^2}{32\pi^2 F_\pi} \;
\epsilon^{\mu\nu\alpha\beta} F_{\mu\nu} F_{\alpha\beta} \pi^0 \,,
\end{equation}
   where $\epsilon_{0123}=1$ and $F_{\mu\nu}$ is the electromagnetic field
strength tensor.
   At ${\cal O}(p^3)$, the LET of VCS is reproduced by the 
tree-level diagrams obtained from Eq.\ (\ref{lpin1pipi2}) and the relevant 
part of the second- and third-order Lagrangian, 
\begin{eqnarray}
\label{lpin2}
\widehat{\cal L}^{(2)}_{\pi N}&=& - \frac{1}{2M} \bar N_v \left[
D \cdot D +\frac{e}{2} (\mu_S+\tau_3\mu_V)
\varepsilon_{\mu \nu \rho \sigma} F^{\mu\nu} v^{\rho} S^{\sigma}\right]
N_v,\\
\label{lpin3}
\widehat{\cal L}^{(3)}_{\pi N}&=&
\frac{ie\varepsilon_{\mu\nu\rho\sigma} F^{\mu\nu}}{8 M^2} \bar N_v
\left[\mu_S-\frac{1}{2}+\tau_3(\mu_V-\frac{1}{2})\right]
S^{\rho} D^{\sigma} N_v +h.c.\,.
\end{eqnarray}

   The linear sigma model (LSM) represents a field-theoretical realization
of chiral $\mbox{SU(2)}_L\times\mbox{SU(2)}_R$ symmetry.
   The dynamical degrees of freedom are given by a nucleon doublet
$\Psi$, a pion triplet $\vec{\pi}$, and a singlet $\sigma$:
\begin{eqnarray}
\label{t:ss:lsm}
{\cal L}_S&=&i\bar{\Psi}\partial\hspace{-.5em}/\Psi 
+\frac{1}{2}\partial_\mu\sigma\partial^\mu\sigma+\frac{1}{2}
\partial_\mu\vec{\pi}\cdot\partial^\mu\vec{\pi}\nonumber\\
&&-g_{\pi N}\bar{\Psi}(\sigma+i\gamma_5\vec{\tau}\cdot\vec{\pi})\Psi
-\frac{\mu^2}{2}(\sigma^2+\vec{\pi}^2)
-\frac{\lambda}{4}(\sigma^2+\vec{\pi}^2)^2,\\ 
\label{t:ss:lsb}
{\cal L}_{s.b.}&=&-c\sigma,
\end{eqnarray}
   where ${\cal L}_{s.b.}$ is a perturbation which explicitly breaks
chiral symmetry.
   With an appropriate choice of parameters ($\mu^2 <0$, $\lambda >0$)
the model reveals spontaneous symmetry breaking, 
$<\!0|\sigma|0\!>=F_\pi=92.4$ MeV.
   The spectrum consists of massless pions, a massive sigma
and nucleons with masses satisfying the  
Goldberger-Treiman relation  $m_N=g_{\pi N} F_\pi$ with $g_A=1$.
   The symmetry breaking of Eq.\ (\ref{t:ss:lsb}) generates
the PCAC relation 
$$\partial^\mu A_\mu^a=F_\pi m^2_\pi \pi^a.$$
   The interaction with the electromagnetic field is introduced via
minimal substitution in Eq.\ (\ref{t:ss:lsm}).
   The generalized polarizabilities have been calculated in the
framework of a one-loop calculation \cite{t:ss:Metz_1996}.

\begin{figure}[ht]
\begin{center}
\epsfig{file=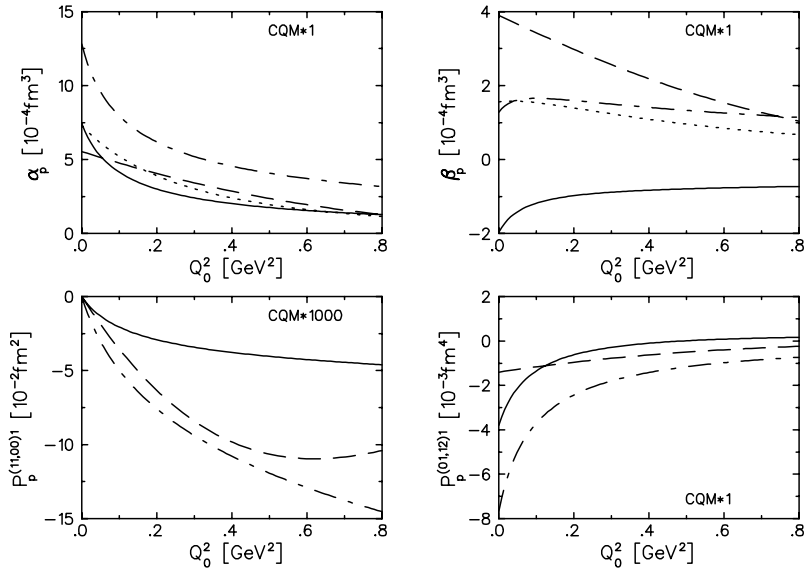,width=300pt}
\caption{Some GPs of the proton. Solid curve: linear sigma model 
\cite{t:ss:Metz_1996}. Dashed curve: Constituent
quark model \cite{t:ss:Guichon_1995}. Dotted curve: Effective
Lagrangian model \cite{t:ss:Vanderhaeghen_1996}.
Dashed-dotted curve: Chiral perturbation theory \cite{t:ss:Hemmert_1997a}.
\label{fig:GPs}}
\end{center}
\end{figure}

   Numerical results for some of the generalized proton polarizabilities 
are shown in Fig.\ \ref{fig:GPs}.
   In both ChPT and the LSM the electric polarizability $\alpha$ decreases
considerably faster with $|\vec{q}|^2$ than in the constituent quark model 
and the effective Lagrangian approach.
   Also, the chiral calculations show a distinct behavior for the
slope of $\beta$ near the origin.
   Note that at ${\cal O}(p^3)$, the results for the GPs are entirely 
given in terms of the pion mass $m_\pi$, the axial coupling constant 
$g_A$, and the pion decay constant $F_\pi$.
   Finally, it has been shown that ChPT and the LSM respect the relations 
between the GPs derived in Ref.\ \cite{t:ss:Drechsel_1997,t:ss:Drechsel_1998}.

\section{Generalized Polarizabilities of Pions}
   At the one-loop level, ${\cal O}(p^4)$, of chiral perturbation 
theory \cite{t:ss:Gasser_1984}, the electromagnetic 
polarizabilities of the charged pion are entirely determined by
an ${\cal O}(p^4)$ counter term \cite{t:ss:Holstein_1990}, 
\begin{equation}
\label{t:ss:alpha}
\alpha=-\beta=\frac{e^2}{4\pi} \frac{2}{m_\pi F^2}(2l^r_5-l^r_6)
=(2.68\pm0.42)\times 10^{-4}\,\mbox{fm}^3,
\end{equation}
   where the linear combination $2l^r_5-l^r_6$ is predicted through
the decay $\pi^+\to e^+\nu_e\gamma$.
   Corrections to this result at ${\cal O}(p^6)$ were shown
to be reasonably small, namely 12\% and 24\% of the ${\cal O}(p^4)$ values
for $\alpha$ and $\beta$, respectively \cite{t:ss:Buergi_1996}.
   
   Presently, the pion VCS reaction is under investigation as part of the
Fermi\-lab E781 SELEX experiment, where a 600 GeV pion beam interacts with 
atomic electrons in nuclear targets \cite{t:ss:Moinester}.    
   In principle, the different behavior under the substitution
$\pi^-\to\pi^+$ of ${\cal M}_{\mbox{\footnotesize BH}}$ and 
${\cal M}_{\mbox{\footnotesize VCS}}$, 
\begin{equation}
{\cal M}_{\mbox{\footnotesize BH}}(\pi^-)= 
-{\cal M}_{\mbox{\footnotesize BH}}(\pi^+),\quad
{\cal M}_{\mbox{\footnotesize VCS}}(\pi^-)
= {\cal M}_{\mbox{\footnotesize VCS}}(\pi^+),
\end{equation}
   may be of use in identifying the contributions due to internal
structure 
by comparing the reactions involving a $\pi^-$ and a $\pi^+$ beam
for the same kinematics:\footnote{This argument works for any particle
which is not its own antiparticle such as the $K^+$ or $K^0$. 
Of course, one could also employ the substitution $e^-\to e^+$.} 
\begin{equation}
d\sigma(\pi^+)-d\sigma(\pi^-)\sim 4 \Re \left(
{\cal M}_{\mbox{\footnotesize BH}}(\pi^+)
{\cal M}^\ast_{\mbox{\footnotesize VCS}}(\pi^+)\right).
\end{equation}
\begin{figure}[ht]
\begin{center}
\epsfig{file=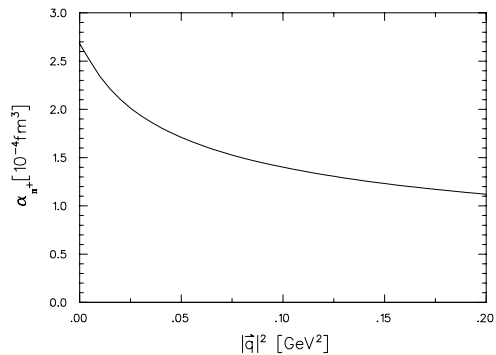,width=130pt,angle=90}
\caption{Generalized polarizability $\alpha$ of the charged pion as
function of $|\vec{q}|^2$ \cite{t:ss:Unkmeir_1998}.
Note that $\alpha(|\vec{q}|^2)=-\beta(|\vec{q}|^2)$ at ${\cal O}(p^4)$.
\label{fig:alphapion}}
\end{center}
\end{figure}
   We have calculated the invariant amplitude for VCS at ${\cal O}(p^4)$
in the framework of chiral perturbation theory.
   The result is in complete agreement with the LET of Ref.\ \
\cite{t:ss:Fearing_1998}.
   Using the procedure developed in Ref.\ \cite{t:ss:Drechsel_1997} we find
for the generalized polarizabilities of the charged pion (see Fig.\
\ref{fig:alphapion})
\begin{eqnarray}
\label{t:ss:genalpha}
\alpha(|\vec{q}|^2)&=&-\beta(|\vec{q}|^2)\nonumber\\
&=&
\frac{e^2}{8\pi m_\pi}\sqrt{\frac{m_\pi}{E_\pi}}
\left[\frac{4(2l^r_5-l^r_6)}{F^2}-2 \frac{m_\pi-E_\pi}{m_\pi}
\frac{1}{(4\pi F)^2} {J^{(0)}}'\left(2\frac{m_\pi-E_\pi}{m_\pi}\right)\right],
\nonumber\\
\end{eqnarray}
where $E_\pi=\sqrt{m_\pi^2+|\vec{q}|^2}$ and 
$${J^{(0)}}'(x)=\frac{1}{x}\left[1-\frac{2}{x\sigma(x)}\ln\left(
\frac{\sigma(x)-1}{\sigma(x)+1}\right)\right],
\quad \sigma(x)=\sqrt{1-\frac{4}{x}},
\quad x\le 0.
$$
   The $|\vec{q}|^2$ dependence does not contain any ${\cal O}(p^4)$ parameter,
i.e., it is entirely given in terms of the pion mass and the pion decay 
constant $F=92.4$ MeV.

\begin{acknowledge}
The author would like to thank D.\ Drechsel, H.W.\ Fearing, T.R.\ Hemmert,
B.R.\ Holstein, G.\ Kn\"ochlein, J.H.\ Koch, A.Yu.\ Korchin, A.I.\ L'vov,
A.\ Metz, and C.\ Unkmeir for a pleasant and fruitful collaboration on
various topics related to virtual Compton scattering. 
   It is pleasure to thank J.M.\ Friedrich, N.\ d'Hose, M.A.\ Moinester,
and A.\ Ocherashvili for useful discussions on experimental issues
in VCS.
\end{acknowledge}

\makeatletter \if@amssymbols%
\clearpage 
\else\relax\fi\makeatother

\SaveFinalPage
\end{document}